\documentclass[journal=jcisd8,manuscript=article,etalmode=truncate, layout=twocolumn]{achemso}

\usepackage{chemformula} 
\usepackage[T1]{fontenc} 
\usepackage{amsmath}
\usepackage{amsfonts}
\usepackage{siunitx}
\usepackage{nameref}
\usepackage{hyperref}
\usepackage{cuted}
\usepackage[absolute]{textpos}

\author{Luisa Vollmer}
\author{Sophie Fellenz}
\author{Fabian Jirasek}
\author{Heike Leitte}
\author{Hans Hasse}
\email{hans.hasse@rptu.de}
\affiliation[RPTU]
{RPTU Kaiserslautern, 67663 Kaiserslautern, Germany}

\title[KnowTD]
  {KnowTD - an actionable knowledge representation system for thermodynamics}

\begin{document}
\begin{textblock}{14.5}(0.5,0.25)
\noindent\footnotesize \textit{This document is the unedited Author’s version of a Submitted Work that was subsequently accepted for publication in the Journal of Chemical Information and Modeling, copyright © 2024 The Authors, published by American Chemical Society after peer review. To access the final edited and published work see} \url{https://pubs.acs.org/doi/full/10.1021/acs.jcim.4c00647}
\end{textblock}
\begin{strip}
    \centering
        \includegraphics[width=0.6\linewidth]{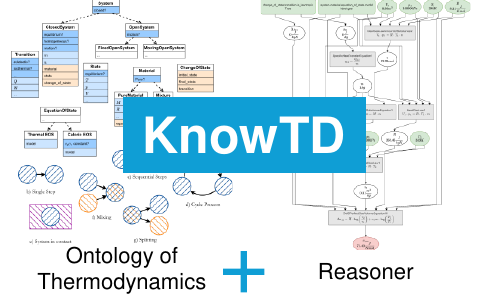}
    \begin{minipage}{0.9\textwidth}
        \begin{abstract}
          We demonstrate that thermodynamic knowledge acquired by humans can be transferred to computers so that the machine can use it to solve thermodynamic problems and produce explainable solutions with a guarantee of correctness. The actionable knowledge representation system that we have created for this purpose is called KnowTD. It is based on an ontology of thermodynamics that represents knowledge of thermodynamic theory, material properties, and thermodynamic problems. The ontology is coupled with a reasoner that sets up the problem to be solved based on user input, extracts the correct, pertinent equations from the ontology, solves the resulting mathematical problem, and returns the solution to the user, together with an explanation of how it was obtained. KnowTD is presently limited to simple thermodynamic problems, similar to those discussed in an introductory course in Engineering Thermodynamics. This covers the basic theory and working principles of thermodynamics. KnowTD is designed in a modular way and is easily extendable.
        \end{abstract}
        \hspace{2cm}
    \end{minipage} 
\end{strip}
\section{Introduction}
\label{sec:introduction}
Thermodynamics is fundamental to both science and engineering and is currently mainly represented in textbooks, lecture notes, and scientific papers. It encompasses some of the most fundamental laws of physics, namely the first and the second laws of thermodynamics, and its applications reach into almost every branch of science and engineering.  Representing thermodynamic knowledge in a consistent way is essential for preserving it, as a basis for its further development, and its applications. 

This requires an appropriate system for representing thermodynamic knowledge - and a system for making that knowledge actionable. Such a system is currently not available. Depending on the problem at hand, applying thermodynamic knowledge may require years of academic training. We do not question the merits of such training (which is an integral part of our daily work), and we are convinced that it will continue to be necessary. However, we are also convinced that routine tasks from thermodynamics can be transferred to machines, allowing well-trained humans to focus on the creative tasks - supported by machines. 

We report here on the development of such an actionable knowledge representation system for thermodynamics, which we call KnowTD. Our approach involves the development of an ontology of thermodynamics and an ontology of the problems the system can solve. It also involves linking the ontology to a reasoning system that solves the problem at hand. 

The work described in this paper provides a starting point for a comprehensive representation of the knowledge in the field of thermodynamics. We start with the basics, as they are taught in introductory courses in Engineering Thermodynamics at universities. The system is designed to make it easily extendable. 

With our work, we address one of the fundamental topics of artificial intelligence: to teach human knowledge to computers. Therefore, this is not a paper on machine learning, it is a paper on machine teaching.  Our ambition is to achieve this for the vast and deep knowledge of thermodynamics. Thermodynamics is particularly suited for this endeavor, as it is a field with a highly developed and well-structured theory, and a successful translation would be particularly rewarding, as thermodynamics belongs to both science and engineering and has extensive applications. The primary aim of the present work is to demonstrate that the significant endeavor of building an actionable knowledge representation of thermodynamics is feasible. It can be assumed that the basic principles that guide the translation of thermodynamic knowledge can also be applied to other fields of science and engineering. 

This paper is structured as follows: we set the stage with some brief remarks on knowledge representation systems. Then, KnowTD is presented, starting with a simple mental model of KnowTD, the ``Lego analogy". Subsequently, the two central parts of KnowTD are presented: the ontology and the reasoner. The ontology is described as a graph with elements of different types (nodes) that stand in relationships (edges). In the reasoner, the problem to be solved is defined, a corresponding system of equations is set up based on the ontology, and the problem is solved. We close with the conclusions and an outlook. The supporting material contains examples for the application of KnowTD and a link to the source code that is provided as a beta version.

\section{Knowledge representation}
\label{sec:knowlege representation}

Knowledge representation for computers is a central topic in artificial intelligence. Know\-ledge representation is concerned with the symbolic representation of knowledge and may also include the automatic reasoning using these representations. 
We use the fundamental concepts of knowledge representation systems as described by \citeauthor{Brachman2004-qz}~\cite{Brachman2004-qz}. These concepts comprise individuals, classes, attributes, relationships, and axioms. \textit{Individuals} are specific instances that we want to reason about, like nitrogen, oxygen, air, or a specific state they may be in. Commonly, these individuals can be grouped into \textit{classes} that feature common characteristics. Such classes include pure substances, mixtures, and ideal gases. A class is a generic prototype for individuals that defines which \textit{attributes} are present and relevant. Attributes assigned to pure substances may be, for example, molar mass, normal boiling point, and critical temperature. While the class only captures the existence of an attribute, the individual contains the precise value, such as oxygen having a molar mass of 16 g/mol. To describe complex systems, we also need relationships and axioms. \textit{Relationships} connect different classes, for instance, materials and their states, and \textit{axioms} define general restrictions, like the mass is never negative.

A \textit{vocabulary} contains the names and definitions of all classes in a list-like structure. Based on this vocabulary, we can begin to think about the organization of the knowledge.
A \textit{taxonomy} additionally includes a hierarchical organization of the classes by relationships, e.g., oxygen is a material. \textit{Ontologies} extend this further and allow arbitrary \textit{relationships} and \textit{axioms}. A variety of methods exist to describe ontologies~\cite{OntologyInCS,Gruber}. Notable examples are~\cite{Mylopoulos1983, 9429985}: logical schemes, network schemes, procedural schemes, and frame-based schemes. Finally, individuals are commonly stored in \textit{databases} that are organized following some schema, i.e., a structural description of the classes.

Any knowledge representation system for a particular domain is based, implicitly or explicitly, on an ontology of that domain. Therefore, in many cases, the development of the knowledge representation system and the development of the corresponding ontology go hand in hand~\cite{OntologyInCS, Gruber, KnowledgeGraphs}. Ontologies can be used in many ways, e.g., to facilitate task specifications, information storage and retrieval (including search), and for structuring the solution of tasks. Ontologies should be targeted towards a specific domain and ideally be interoperable.
Using ontologies has many benefits:  among others, they describe a common vocabulary, promote explication of what has been left implicit, enhance systematic procedures, enable standardization, and provide meta-model functionality~\cite{OntologicalEngineering}. 

While, to the best of our knowledge, no ontologies (or knowledge representation systems) exist that are explicitly dedicated to thermodynamics, there are ontologies of related fields, and there are also ontologies that include some thermodynamic topics. PhysSys~\cite{PhysSys} provides a formal conceptual foundation of computer-aided engineering systems with a focus on creating a common understanding between design engineers and end users. EMMO~\cite{EMMO} is an ontology developed by the European Materials Modelling Council (EMMC) to provide a common semantic framework for describing materials, material models, and data on materials. ChEBI~\cite{ChEBI} (Chemical Entities of Biological Interest) is a database and ontology for chemical entities related to biology, in which an ontology is used to structure the database and define logical relationships. The Ontology of Physics for Biology (OPB)~\cite{OPB} was designed for annotating biophysical properties in databases and analytical models. 
Some ontologies are from the chemical engineering domain and contain thermodynamic elements: OntoCape~\cite{OntoCape} provides a general framework for computer-aided process modeling. OntoKin~\cite{OntoKin} is an ontology that covers reaction mechanisms and reaction kinetics and includes semantics for describing chemical reaction networks. Propnet~\cite{propnet} uses connected physical relationships to enhance materials property data, aiding in the discovery of correlations and the design of multifunctional materials. 

Ontologies that focus on procedural issues include: OSMO~\cite{OSMO}, the Ontology for Simulation, Modelling, and Optimization, which focuses on workflows in computational molecular engineering and material modeling, and PSO~\cite{PSO}, the Physics-based Simulation Ontology, which addresses similar issues on a more general level. There are also meta-ontologies that aim at combining ontologies from different fields, such as the World Avatar \cite{QAChemistry}, which aims at describing and simulating complex processes \cite{DigitalTwin}. 

Thermodynamic knowledge also includes the knowledge of thermodynamic properties of materials. This information is typically stored in electronic databases. For example, there are large databases that aim to collect all thermodynamic data of fluid materials that have ever been published and to represent that data in a structured way after excluding obviously faulty data points: the Dortmund Data Bank~\cite{DDB} and the DIPPR Data Base~\cite{Thomson1996}. In general terms, these are databases in which information on individuals from the class fluid materials are stored.

\section{Making knowledge actionable}
\label{sec:Action}

Having discussed ways in which knowledge can be stored, we can now focus on the application of this knowledge to particular tasks.
In general, knowledge representation systems can serve a variety of tasks, including problem solving, knowledge retrieval, and educational use. 

In mathematics, multiple problem-solving strategies have been developed that combine ontologies with reasoning: MathGraph \cite{MathGraph} is an equation-based system that focuses on analytical problems, but there are also systems for solving discrete problems \cite{OntologyProblemSolver, SolidGeometry} and geometric problems \cite{Kurbatov}. In the latter, the solution and its development are represented graphically. Combining ontologies and reasoning has the advantage of obtaining explainable solutions. It can be tracked which parts of the knowledge were used to obtain the solution and how they were used. The way solutions are obtained is similar to that used by humans \cite{OntologyProblemSolver}. Therefore, the results will be more readily accepted than those of less transparent methods. 

Ontologies are also used in question-answering systems. Thereby, the relationships between the concepts defined in the ontology are used to answer questions. If, e.g., the ontology connects the concept ``temperature''  to the concepts ``Kelvin'', ``Celsius'', and ``Fahrenheit'' by a ``is unit of''-relationships, the question-answering system can answer questions like "What is the unit of the temperature?”. Examples of such systems are QAPD \cite{QAPD} (a question-answering system in the physics domain), and the question-answering system implemented on top of The World Avatar \cite{QAChemistry}. 

\section{KnowTD}
\label{sec:KnowTD}

\subsection{Overview}
\label{sec:KnowTD/Overview}
\begin{figure*}[h]
    \centering
    \includegraphics[width=.9\linewidth]{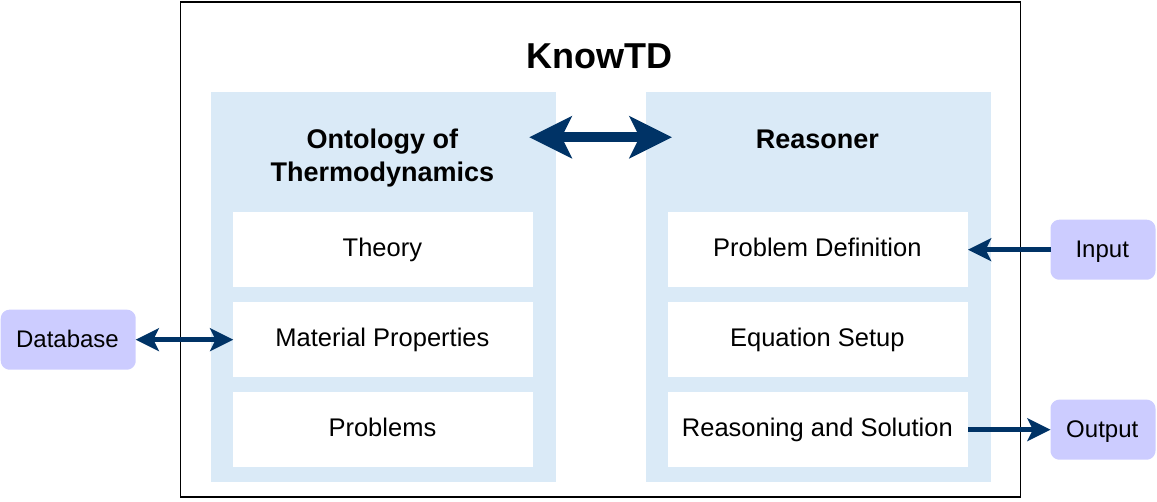}
    \caption{Scheme of the actionable knowledge representation system KnowTD, including the ontology and the reasoner.}
    \label{fig:system}
\end{figure*}

To develop a system for the actionable representation of thermodynamic knowledge, two key tasks need to be solved:
i) find an appropriate representation of thermodynamic knowledge and thermodynamic problems and ii) devise a strategy to effectively use this knowledge to solve actual problems.
To accomplish this, KnowTD has two main components: i) an ontology of thermodynamics and ii) a reasoner. An overview of the structure of KnowTD is given in Fig.~\ref{fig:system}.

There are two main components of KnowTD: the ontology and the reasoner (see Fig \ref{fig:system}). The ontology of thermodynamics is a digital repository of thermodynamic knowledge and encompasses the thermodynamic theory, the knowledge of material properties (which may also be taken from external databases), and the knowledge of the structure of thermodynamic problems, cf. Fig~\ref{fig:system}. The reasoner uses the knowledge stored in the ontology to solve problems specified by the user input. This is done in a sequence of three steps: the problem definition, the set-up of the system of equations, and the reasoning, which includes solving the problem. 
In its current version, the reasoner simply checks if, based on the equations that describe the problem, a feasible solution path can be found - assuming that a sequential, step-by-step solution is possible. More sophisticated schemes to analyze and solve the equations should be implemented in future work, in which the reasoner could also be extended to identify ill-posed problems and/or request additional information to ensure a well-defined and solvable problem. 

Currently, KnowTD is limited to functionalities needed to solve thermodynamic problems involving reversible and irreversible changes of state in homogeneous closed systems containing ideal gases. However, this already comprises many constitutive parts of thermodynamics, such as the first and second law, knowledge of thermal and caloric properties of materials, etc. The current version of KnowTD is only a starting point, which is why we have taken care to keep it extendable. As it stands, KnowTD also provides a proof of principle: it shows that useful, actionable knowledge representation systems for thermodynamics can be built. 

While we focus here on problem-solving, we emphasize that KnowTD is also attractive for other fields, e.g., academic education: the goal of academic education is to enable students to solve problems - while the goal of KnowTD is to enable machines to solve problems. It is highly interesting to compare the approaches in both fields.

\subsection{A mental model: the Lego\textsuperscript{\textregistered}-analogy}
\label{subsec:Lego}

For understanding the conceptual framework of KnowTD, it is helpful to establish a simple mental model, which we call the \textit{Lego\textsuperscript{\textregistered}-analogy}, referring to the famous brick toy.  The ontology of thermodynamics uses fundamental building blocks akin to Lego bricks. Examples are: closed system, material, pressure, temperature, state, change of state, heat, work, 1st law, 2nd law, thermal equation of state, and caloric equation of state. Just as Lego bricks must be joined together to build a model of a given object, e.g., of a house, the elements from the ontology must be joined together to build a thermodynamic model, e.g., of a compression of a gas in a cylinder.  
KnowTD is a highly flexible system that can handle a large variety of tasks and create corresponding thermodynamic models. KnowTD does not rely on a database of pre-configured models. Instead, it \textit{knows} how to build models from the different types of blocks.

In the building process, the connection of the elements plays a key role. While Lego bricks have only a single type of connector, the elements of the ontology stand in different types of relationships. Furthermore, with Lego, one can always choose whether or not to use a connecting option, while in the ontology of thermodynamics, there are mandatory connections (e.g., a system needs to have a material). These mandatory connections result in what one can imagine as blueprints of fixed units of Lego bricks that are used to create the model.   

The Lego-analogy illustrates an essential point of thermodynamic theory: complex structures can be constructed from a limited set of elements. The number of basic concepts in thermodynamics is not large, but their judicious combination makes it possible to solve a huge variety of tasks. 
Mastering how to assemble these elements effectively to obtain the desired object presents a challenge which we explore in subsequent chapters. The Lego-analogy also highlights the possibility of discovering missing parts while working on a task and the possibility of encountering unsolvable tasks.

So far, we have only considered the building of the thermodynamic model, but KnowTD continues beyond that. To stay in the analogy: once the Lego-model is set up, one can start playing with it. In KnowTD, this means carrying out the actions the model was designed for, e.g., solving problems for different sets of input parameters. In principle, this is similar to working with a standard process simulator; however, it has some noteworthy special features, the most important of which is that the entire theory behind the solutions is always traceable and can be fully documented, if required. 

\subsection{Current scope of KnowTD}
\label{subsec:Scope}

Given the vast amount and complexity of thermodynamic knowledge, creating a knowledge representation system for thermodynamics is an extremely ambitious endeavor that has to be tackled step by step. The scope of the first step, on which we report here, was directed to those parts of thermodynamics that are relevant for solving problems related to a change of state of a closed system containing an ideal gas. It covers the essential elements of thermodynamics, including the first and second laws as well as the interplay between theory and material properties - so that it is well-suited for our present purposes. The choice of this topical area is also motivated by the fact that it is usually chosen as the starting point of academic courses in Engineering Thermodynamics. Furthermore, the main task of the present research was to find a suitable structure for the knowledge representation system, which does not necessarily require a broad scope if it covers the key issues and if the system is designed to be extendable. 

\subsection{Ontology}
\label{sec:Ontology}

The goal of our ontology of thermodynamics is twofold: first, to provide a knowledge representation that is accessible to both humans and computers; and second, to enable computer programs to reason about thermodynamic problems and solve them - with a guarantee that the solution is correct.

\subsubsection{Elements}
\label{subsec:Elements}

In an initial step, we have compiled 160 key terms used to describe thermodynamic problems and their solutions within the scope defined in section~\textit{\nameref{subsec:Scope}}. These terms are listed in electronic form in the Supporting Information. We do not claim that this list is complete, but we found it sufficient for the purposes of the present work. The list contains redundant terms, but only if all versions are widely used.  

Each term is considered an element of the ontology, was assigned a unique name, and cross-referenced with synonyms to enhance clarity and avoid ambiguity. 
This collection of terms forms a vocabulary that comprises well-defined terms for the various entities within the ontology. The vocabulary can also be interpreted as a list of named elements from which the ontology can be built. How this was done is explained in the following sections. This procedure is also a blueprint for future extensions of KnowTD.

We have refrained here from giving formal definitions of the various terms but have added comments wherever misunderstandings might arise; cf. Supporting Information. These comments can be considered as attributes of the elements of the ontology. We plan to extend this later with formal definitions.
For now, we refer to textbooks on thermodynamics for such definitions, e.g., to~\citenum{osti_6250731, smith2018introduction, Grigull1977}~\cite{osti_6250731, smith2018introduction, Grigull1977}. 

\subsubsection{Basic classifications}
\label{subsec:Basic classifications}

The basic elements from section~\textit{\nameref{subsec:Elements}} were assigned to coherent classes and then organized in a hierarchical structure. This describes abstract relationships and thus facilitates understanding. It also lays the foundation for working with the elements. 

All elements were assigned to one of the five following classes:
\begin{itemize}
    \item Concepts ($C$)
    \item Variables ($V$)
    \item Attributes ($A$) 
    \item Equations ($E$)
    \item Rules ($R$).
  \end{itemize}

 Details are described in the Supporting Information. 
 
The concepts $C$ are basic elements of the thermodynamic theory, such as: thermodynamic system, closed system, state, material, equation of state, change of state, etc. The variables are used for the mathematical description of thermodynamic problems (e.g., mass $m$, temperature $T$, pressure $p$, volume $V$), and equations relate variables (e.g., the ideal gas law relates $m, p, V$, and $T$).  Attributes describe non-numeric characteristics of a concept. A process, for example, can be adiabatic, isobaric, isothermal, etc. Attributes are vital for deciding which part of the knowledge is applicable. If a process is adiabatic, then there is no heat transfer ($Q = 0$). The corresponding if-then relationships are expressed in the ontology using rules. KnowTD primarily uses rules to choose applicable parts of the knowledge according to the specification of attributes. This can also involve a combination of attributes, e.g., if a process is adiabatic and reversible, it is isentropic and $\Delta s = 0$.
We emphasize that the ontology describes classes of objects and their relations, while instances of these classes are only defined upon applying the ontology and must obey the rules formulated for the classes. 

\subsubsection{Relationships and hierarchy}
\label{subsec:Relationships}

To create a useful ontology, not only the elements but also their relationships must be described. We will start here by discussing the relationships between the different concepts and then continue with those between the concepts and the variables. The relationships between the equations and the variables are business-as-usual in mathematical modeling and are therefore not discussed here. The relationships between the concepts and the equations are essential for making the ontology actionable and are described separately in the next section. 

We distinguish two major classes of relationships:
\begin{itemize}
    \item ``\textit{has\_a}" relationship
    \item ``\textit{is\_a}" relationship
\end{itemize}
In KnowTD, \textit{has\_a} relationships are used to link concepts to other concepts ($C$) and their respective attributes ($A$) and variables ($V$). Here are some examples for \textit{has\_a} relationships:
\begin{itemize}
    \item Closed system ($C$) \textit{has\_a}  material ($C$)
    \item Material ($C$) \textit{has\_a} thermal equation of state ($C$)
    \item Closed system ($C$) \textit{has\_a}  state ($C$)
    \item Closed system ($C$) \textit{has\_a}  mass ($V$)
    \item Closed system ($C$) \textit{has\_a}  homogenous? ($A$)
\end{itemize}

In the last example, the attribute "homogenous?" ($A$) refers to a property of the closed system that indicates whether the system is homogenous or heterogeneous. 

In an ontology, all relationships need unique names, i.e., the complete names of the relationships above would be $has\_a\_material$, $has\_a\_state$, etc. These long names ensure clear communication but also lead to verbose descriptions. For clarity of the writing, we have shortened these long names in this manuscript.

The \textit{is\_a} relationships define specializations of elements in the ontology, i.e., we can express that an element is a specialized version of a broader element. Some examples are:
\begin{itemize}
    \item Closed system (C) \textit{is\_a} system (C)
    \item PureMaterial (C) \textit{is\_a} material (C)
    \item Mixture (C) \textit{is\_a} material (C)
\end{itemize}
The \textit{is\_a} relationships help to structure large corpora of knowledge as they induce a hierarchy from broad general elements to fine-grained specialized ones. Which of the fine-grained elements applies in a given problem is controlled by attributes of the parent element.
Specialized elements inherit all \textit{has\_a} relationships from their parent and may extend them with additional relationships and rules. 

The hierarchical structuring of the elements in the ontology enables the flexible usage of the knowledge in the reasoning phase. We can specify on a very general level that any system must have a material. This requirement then also holds for any specialized types of systems, e.g., for a ``closed system". We will make extensive use of this hierarchical structuring in the reasoning phase when we solve actual problems, see section~\textit{\nameref{sec:Reasoner}}.

\subsubsection{Representing KnowTD as a knowledge graph}
\label{subsec:KnowledgeGraph}

The two most common ways to represent complex knowledge are representations inspired by file browsers and graph-based representations. Both ways are supported, e.g., by ontology editors such as Protégé.  We use the graph-based approach here and a notation similar to class diagrams.

\begin{figure*}
    \centering
    \includegraphics[width=.8\textwidth]{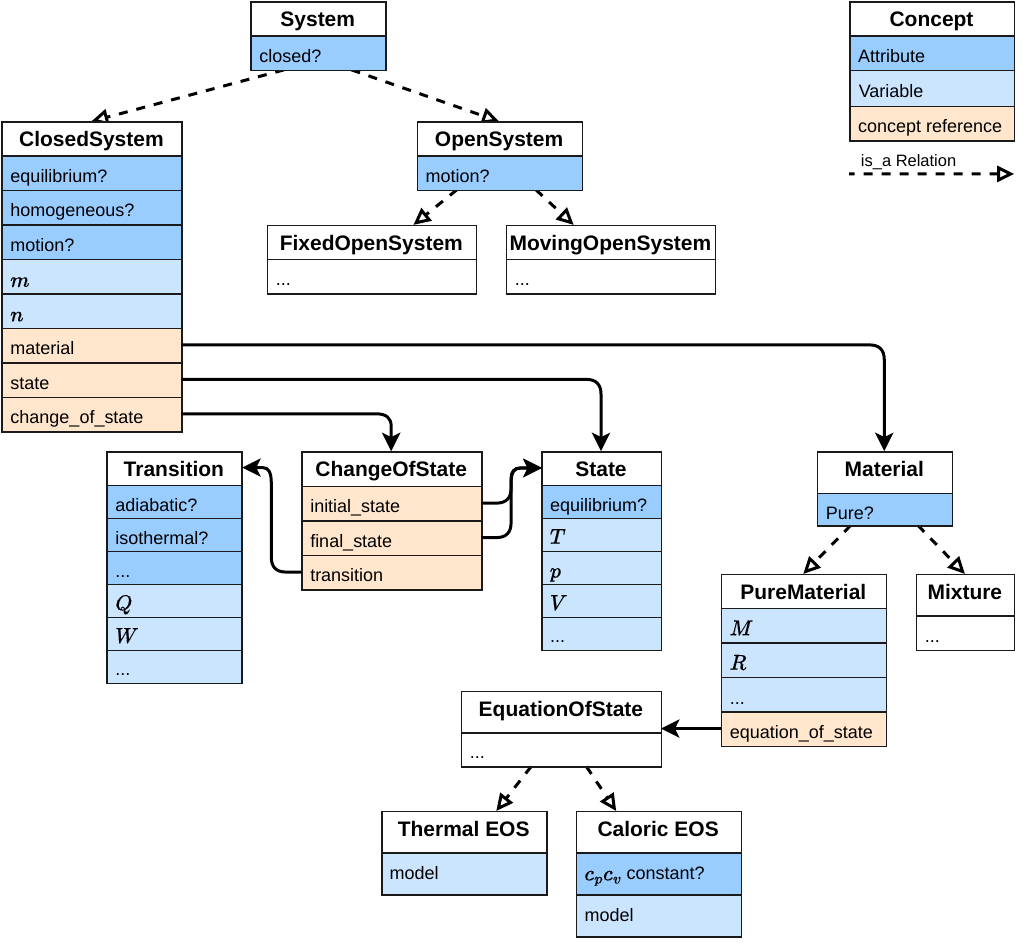}
    \caption{Subgraph of the current KnowTD knowledge graph showing some important concepts, variables, and attributes and their relationships. Dashed arrows correspond to \textit{is\_a} relationships. The list-like entries correspond to \textit{has\_a} relationships, orange entries in the list refer to a concept. The shown attributes control which of the \textit{is\_a} relationships applies.}
    \label{fig:class_diagram}
\end{figure*}

Even in its present, preliminary form, the knowledge graph of KnowTD is so complex that it cannot be displayed as a whole in an informative way. Filtered views are more useful. An example of such a subgraph is shown in Fig.~\ref{fig:class_diagram}, in which some important concepts and their relationships are depicted. Also, some examples of relationships of these concepts with variables are included; equations are not shown in Fig.~\ref{fig:class_diagram} for clarity. The boxes represent elements (concepts: white and orange; variables blue). The \textit{has\_a} relationships are indicated by white boxes heading a list of orange or blue boxes (meaning: "white \textit{has\_a} orange or blue"); solid arrows indicate that a sequence of \textit{has\_a} relationships exists. The \textit{is\_a} relationships are indicated by dashed arrows. $A$ $\rightarrow$ (dashed) B reads: $B$ \textit{is\_a} $A$. 
To illustrate that the approach can be extended, some elements not yet implemented are shown in Fig.~\ref{fig:class_diagram}. The full information on the relationships between the elements of the ontology that are currently considered in KnowTD is given in electronic form in the Supporting Information.

\subsubsection{Equations in the ontology}
\label{sec:equations}
So far, we have discussed how to represent the conceptual aspects of thermodynamics know\-ledge. In order to create mathematical models and actually solve thermodynamic problems, equations that describe the behavior of thermodynamic systems mathematically need to be included in KnowTD.

Equations (E) relate variables (V). The relations in which they stand can be represented as edges of a bipartite graph~\cite{pothen1990computing}. Equations and variables are the nodes of the graph. An equation node is linked to a variable node if the equation contains the variable. Bipartite means that the nodes can be grouped into two sets such that there are no edges connecting nodes within the same set. In our case, the two groups are equation nodes and variables nodes. 

Furthermore, all variables are assigned to concepts by \textit{has\_a} relationships. Thereby, an equation may link variables that are assigned to different concepts; e.g., applied to a closed system, the ideal gas law ($p \cdot V = m \cdot R \cdot T)$ links variables belonging to that system ($m$) with its material ($R$) in a given state ($p, V, T$). To handle this, we store all equations in an augmented machine-readable form. For the ideal gas law, this reads $p_{\text{state}} \cdot V_{\text{state}} = m_{\text{closed system}} \cdot R_{\text{material}} \cdot T_{\text{state}}$ indicating for each variable to which concept it belongs (compare also Fig.~\ref{fig:class_diagram}.
The explicit assignment of the variables to concepts is important for the application of the equation: in our example, the ideal gas law might apply for different states (with different values of $p, V, T$) of a given system with a given material, i.e., with the same values of $m$ and $R$.

To decide whether an equation is actually applicable (i.e., whether the relation between variables it represents is valid in a given situation), rules (R) may be needed, which can be assigned to any equation.

The equations presently implemented in KnowTD are listed in the Supporting Information. They are mostly linear algebraic equations, but there are also non-linear equations, which mainly stem from the description of material properties (i.e., the ideal gas law or the logarithmic terms occurring in models of the entropy). Some of the equations are redundant, as the omission of any redundancy would have resulted in the elimination of many useful equations. 

\subsubsection{Representation of thermodynamic processes}\label{sec:ProblemRepresentation}
\begin{figure*} [t]
    \centering
    \includegraphics[width=0.8\textwidth]{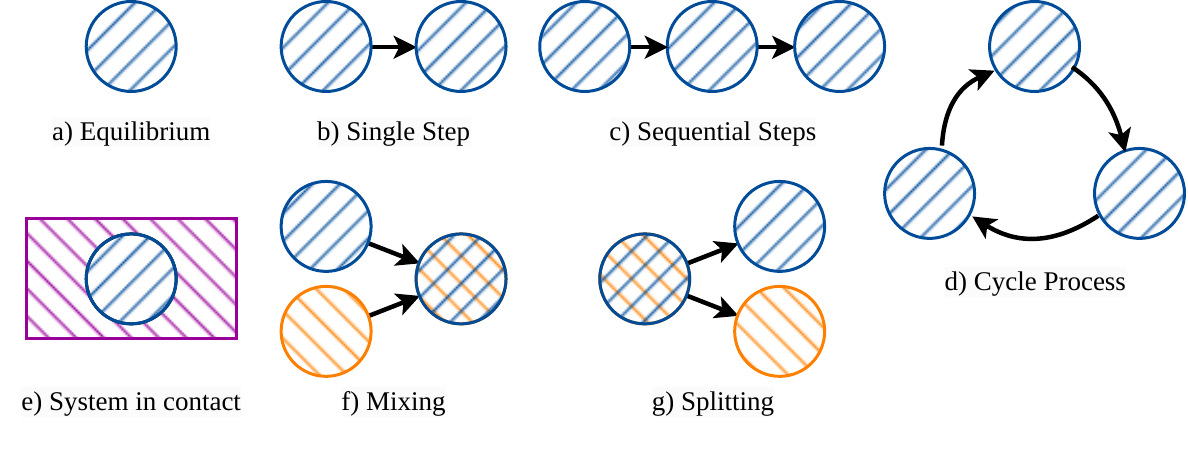}
    \caption{Process classes of closed systems.}
    \label{fig:Processes_with_systems}
\end{figure*}

To solve thermodynamic problems, it is not sufficient to know thermodynamic theory alone; it is also essential to have knowledge of the process under scrutiny. Fig.~\ref{fig:Processes_with_systems} shows sketches of some elementary classes of processes with closed systems. The basic elements from Fig.~\ref{fig:Processes_with_systems} can be combined to describe more complex processes, i.e., they can be considered building blocks of process models. The process classes are stored in the ontology of problems (cf. Fig.~\ref{fig:system}). They serve as a template; depending on the selected class, corresponding instances of the concepts are created and the given values from the task are assigned in the reasoner.

The most simple building block is the system that undergoes no changes, i.e., a system in equilibrium, cf. Fig.~\ref{fig:Processes_with_systems} Panel a); the most elementary process is the single-step process shown in Panel b). KnowTD currently covers these two classes but is designed to be extendable to other types of processes. 
This is straightforward for the sequential multi-step process depicted in Panel c) and the cyclic process from panel d). They can be modeled by different combinations of process classes a) and b), eventually using additional class-specific concepts. Other extensions depicted in Fig.~\ref{fig:Processes_with_systems} are: Panel e): systems in contact, which includes the consideration of a system in contact with its environment; Panel f): merging two systems (mixing); Panel g): splitting a given system. Fig.~\ref{fig:Processes_with_systems} also illustrates that thermodynamic processes can be represented by graph-like structures.

\subsubsection{Implementation of the ontology}
\label{subsec:Implementation}

In the previous sections, we have described the theoretical framework of the ontology of thermodynamics. To make it actionable, it has to be implemented on computers. For ontology implementation, several standardized formats exist, including the Resource Description Framework (RDF)~\cite{RDFresource}, the Web Ontology Language (OWL)~\cite{OWL}, and the Open Biomedical Ontologies (OBO)~\cite{OBO}. 

We have implemented the ontology using the LinkML schema language. LinkML~\cite{LinkML} is a general-purpose modeling language designed for working with linked data. Model descriptions in LinkML, so-called schemas, are provided in YAML file format. YAML files use a minimal amount of markup (text elements that help the computer understand the file) and look very much like formatted lists in text documents.

The reason for choosing LinkML was that it has many convenient features that make it easy to develop and work with the ontology. Firstly, LinkML has a rich set of converters. The thermodynamics ontology written in a LinkML schema can thus be converted easily to several standard formats, including OWL, python data classes, and markdown. OWL, a standard file format for semantic web ontologies~\cite{OWL}, supports integration with editors such as Protégé for building and maintaining ontologies~\cite{Protege}. The python representation not only enables ontology editing but also allows to apply python code to work with the elements defined in the ontology. The markdown representation can be used to generate an interactive documentation of the ontology. 

Additionally, LinkML provides a validator. The validator is a computer program that accepts a schema definition (the ontology of thermodynamics) and a data file (the problem definition) and checks that the data file conforms to the schema. 

\subsection{Reasoner}
\label{sec:Reasoner}
In the previous section, we have explained how thermodynamic knowledge, including knowledge of thermodynamic processes, can be represented in an ontology.  In the present section, we explain how the knowledge stored in the ontology can be used to solve thermodynamic problems. The corresponding tasks are solved in the reasoner, cf. Fig.~\ref{fig:system}.

The reasoner operates in three steps (cf. Fig.~\ref{fig:system}): 

I) \textit{Problem Definition}: Here, the user specifies the type of problem to be solved and provides the input data. KnowTD then creates a conceptual model of the problem. For example, if a simple change of state is considered (cf. Fig.~\ref{fig:Processes_with_systems} Panel b), the conceptual model comprises that two equilibrium states of a given closed system have to be considered that are connected by a single change of state. A list of corresponding variables is created and values are assigned if available from the input. Also, target variables are defined.

II) \textit{Equation Setup}: In this step, KnowTD collects all applicable equations for the specified problem. The resulting equations are all valid, but there will usually be redundant equations that are not needed. 

III) \textit{Reasoning and Solution}: In the last step, KnowTD analyzes the set of equations from Step II and computes the numerical solution to the given problem. While the equation setup provides all related equations potentially relevant to the solution, Step III filters the set to those that actually contribute to the solution.

The outcome goes beyond providing a solution by giving just the numbers for specific variables of interest; it also includes a detailed, step-by-step report on how that solution was derived.

In the following, the three steps are discussed in more detail. 

\subsubsection{Problem definition}
\label{subsubsec:InstanceDef}
\begin{figure*} [ht]
    \centering
    \includegraphics[width=.8\textwidth]{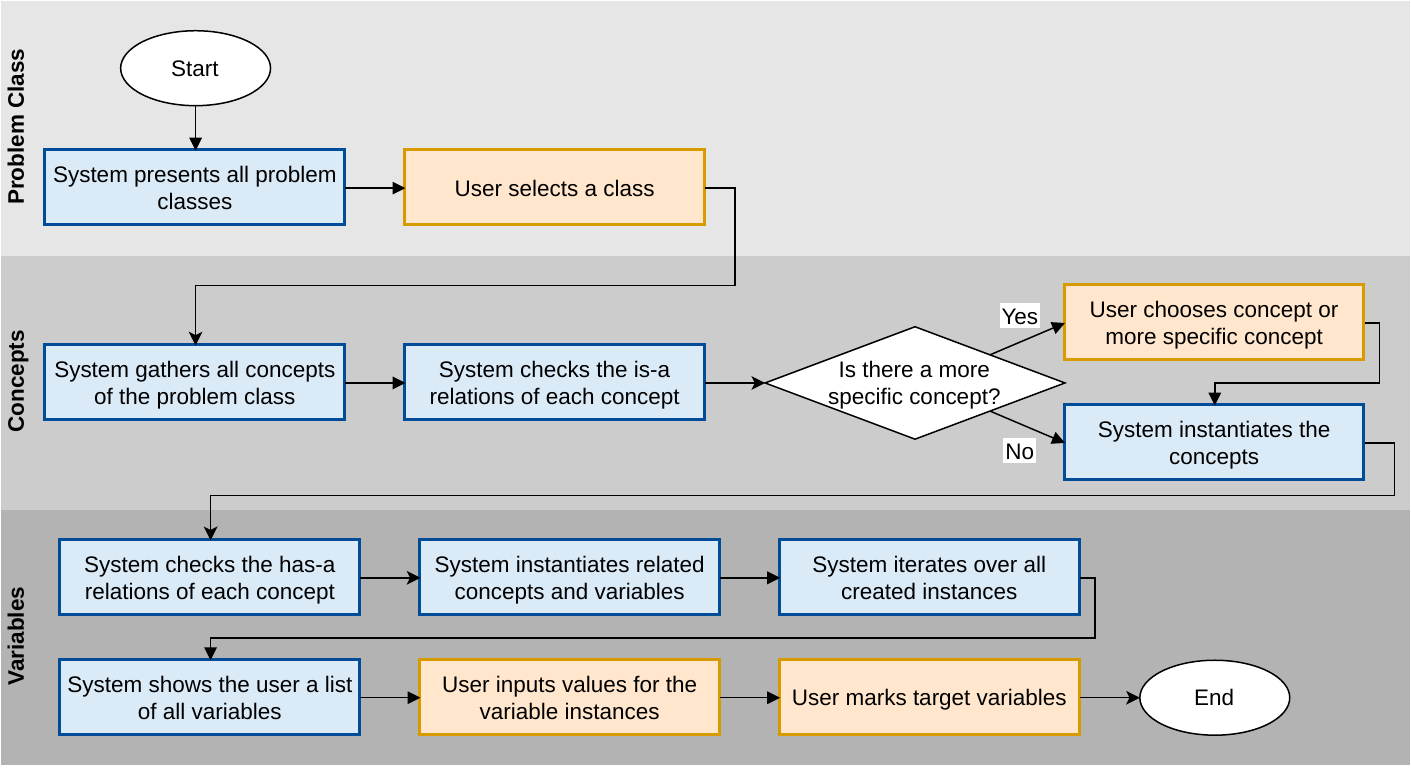}
    \caption{User interaction diagram of KnowTD. The system assists the user in translating the given problem to the input of KnowTD. It uses the knowledge of the ontology to create instances of the concepts.}
    \label{fig:user_dialogue}
\end{figure*}

The goal of the first step is to create a conceptual model of the given problem and to identify the parts of the ontology that are relevant to its solution. To achieve this, the user is guided through an interactive dialogue. An overview of the steps in the dialogue is given in Fig.~\ref{fig:user_dialogue}.

First, the user has to select a specific process class. This allows KnowTD to select appropriate base concepts (C) from the ontology, i.e., it can initialize a closed system, material, states, and changes of state, as appropriate for the selected problem. These base concepts may be related to other concepts by \textit{is\_a} relationships that open choices. The user gets a list of possible choices related to each concept and can select the appropriate one by setting the corresponding attributes (A). The corresponding concept is then instantiated. 

For all instantiated concepts, KnowTD selects all variables related to the concept by a \textit{has\_a} relationship and presents an according list of variables to the user. The names of the variables are generated automatically using default values, e.g., if a closed system undergoes only one change of state, the temperature for the initial state is named $T_1$ and that for the final state $T_2$. All names can be altered by the user.

In the input dialogue, the user may then enter values for any of the pertinent variables. For variables that have dimensions, the input is made in SI units. The corresponding SI unit is defined in the variable node of the ontology. Other units need to be converted before entering the input. Besides setting the values for these input variables, the user may also select target variables for which results are to be computed. If no target values are selected, KnowTD will consider all variables that are instantiated and are no input variables as targets and will try to assign values to all of them.

There are also some variables for which values are always set. In the current version, these are the universal gas constant 
$R = \SI{8.31446261815324}{\joule\per\mol\per\kelvin}$ and the reference state used for the normalization of caloric and entropic properties $T_0 = \SI{293.15}{\kelvin}$ and $p_0 = \SI{0.1}{\MPa}$.

Thermodynamic problems are typically given in text form. The link between such text and the formatted input that KnowTD requires currently has to be made by a user. Thereby, after having identified the task class, the user simply has to decide whether or not information on a certain variable from a list is available, and if it is available, the given value has to be entered. This is a much simpler task than solving the actual problem. Nevertheless, enabling KnowTD to work directly with text would be highly desirable.  The present work lays the foundation for this by providing straightforward questions for which the answers have to be retrieved from the text, which, in turn, is a feasible task for large language models.

\subsubsection{Finding the mathematical equations that describe the problem}
The final goal of KnowTD is to solve the given problem and present an interpretable solution with a guarantee of correctness and a step-by-step explanation of how it was derived. So far, we have characterized the problem and identified the parts of the ontology relevant to its solution. The next step is to set up the corresponding mathematical equations that describe the problem and can be used to solve it. From the previous steps, we know all the variables in our problem and have already given them unique names. The ontology also contains a list of equations that relate the variables. Thereby, only equations are considered for which all rules that are connected to the equation are fulfilled. In the extended notation that we have used for the equations, each variable is linked to a specific concept. This enables the creation of instances of the equations based on the instances of the concepts. This leads to a valid system of mathematical equations that describe the problem at hand. However, that system generally contains redundant equations, which makes it different from a typical mathematical model of the considered processes.  

Methods for eliminating redundant equations from sets of linear equations exist, but we are not aware of any general method that could accomplish this for the non-linear sets of equations that we have to deal with here. However, including valid redundant equations in a set of equations does not prevent the solution, so we have decided to work with the full set of valid equations for the purposes of the present work. This is done in the reasoning step.

\subsubsection{Reasoning}

The task of the reasoning step is to find the correct values for the target variables for the given input variables and the given system of equations, which is set up as described in the previous paragraph.  Unfortunately, to the best of our knowledge,  a robust and generally applicable method to accomplish this for the systems of non-linear equations in our prototype does not exist. We have, therefore, developed an ad hoc approach that can solve many, but not all, of the problems at hand. The development of a more rigorous and comprehensive approach to solving this problem is left open for future work.  

Before we explain our approach, let us briefly discuss how a human expert would solve problems like the ones we are considering here. He would probably not start by writing down all the relevant equations, as we have done. Instead, he would start with some of the given variables and try to find a path to some of the target variables by applying relevant equations he knows and then somehow extend this to the full set of target variables, using the full information about the input variables. Strategies for doing this are taught in schools and universities, but for the most part, they are not explicit knowledge. This intuitive approach inspires the method we have used here to solve the problem which can be seen as a formalization of it. 

As already pointed out in the section~\textit{\nameref{sec:equations}}, the equations and variables in a given problem can be represented in a bipartite graph. Our reasoning algorithm operates on this graph, which we call the reasoning graph. The graph is initialized by creating nodes for all equations and variables identified during the previous steps. Edges link equations and their respective variables. Our approach is basically a brute force solution that is inspired by the Dulmage-Mendelsohn decomposition~\cite{pothen1990computing,dennis1994triangular}. It operates in two steps: (i) We first solve a reachability problem, i.e., find all variables in the reasoning graph that can be computed given the input. (ii) In the second step, we search for an efficient path that leads from input to output variables. 

The approach is currently limited by the following restriction: we require that the equations be solved sequentially and only one at a time. If an equation has five variables, such as the ideal gas law, we require that four of them are known and then assume that the fifth can be calculated. This leads to what we call a \textit{single equation traversal} of the reasoning graph. This is a harsh restriction; even in simple cases, solutions might require, e.g., substitution or the solution a system of equations. There are many different approaches to tackle such cases, so many that we refrain from entering a discussion here. We simply emphasize that the restrictions coming from the single equation traversal are not fundamental; this approach should be considered as an example of how solutions to the mathematical problem at hand can be obtained.

\paragraph{Graph-based reachability analysis} The reasoning graph created in the previous steps is an undirected graph. In the present step, it is turned into a directed graph by encoding how available information can be used to derive new pieces of information. In the beginning, only the given variables are known. Hence, we turn all edges associated with their nodes into outgoing edges, i.e., they can share their information with equations that use these variables. Now, we iteratively check all equations in the graph to see if they only have one unknown variable left, i.e., for an equation node of degree k, if it has k-1 ingoing edges. If so, the unknown variable can be computed, and we turn the last edge into a directed one going out from the considered equation node to a formerly unknown variable $x$. As $x$ is now known, we change all its other edges (links to equations that also use $x$) into outgoing edges, i.e., the equations can use the value computed for $x$. This procedure is repeated until no additional information can be computed. If all target values have at least one ingoing edge, the problem can be solved by single equation traversal.

\paragraph{Retrieving the Solution Path} To obtain the solution path from input to output, we backtrack the information in the Reasoning Graph. Therefore, we compute the ancestors of all output nodes. This may yield different options to solve the problem. From these, any path may be chosen. Criteria for this could be to choose the path with the smallest number of equations or to choose paths that avoid equations that can only be solved numerically. We have not explored any of these options in the present work and leave this open for future work. Here, we have simply taken the first path that was found in the reachability analysis by which all target variables were reached. The solution is provided in SI units; should other units be required, the user must convert the results. As the conversion of units is a solved problem, an automatic conversion may be integrated into the user dialogue in future work.

\paragraph{Remarks on the reasoning procedure} The current reasoning procedure is just a first simple approach to tackle the mathematical problem at hand and is limited by the requirement of a single equation traversal. Many other options exist and wait to be explored. If only one solution path by single equation traversal exists, the algorithm will find it. If several paths exist, the algorithm will find one, but it could be extended straightforwardly to determine all solution paths. Our algorithm can efficiently handle systems of equations with redundancy, which underpins that there is no need to find a minimal set of equations. Here, we have considered only mathematical problems with a unique solution (in the sense that there is only a single set of target variables for the given set of input variables that fulfill the equations). This is typically the case for the problems that we consider here if a solution exists at all. We point out that the reasoning procedure does not rely on any domain knowledge and can in principle be applied to any domain that requires to set up and solve a system of equations.

\subsection{Implementation of the reasoner}
The KnowTD reasoner was implemented using Python. It can easily access the information from the ontology as the ontology is implemented using LinkML, which offers a conversion to Python code.   
LinkML also provides a conversion of schema definitions to python data classes. Thus, it automatically generates a python package to import data into python objects meaning that we did not have to write extra code to load problems into a python program. Python also provides a variety of packages to support the reasoning tasks. For the graph implementation and analysis, we have used the networkx package and for symbolic reasoning sympy. In the prototype, no results from the reasoner are cached, but this is an interesting option for an extension and would, e.g., significantly facilitate solving tasks that require parametric studies.

\section{Case studies and prototype}
\label{sec:Tests}

A set of 13 thermodynamic problems from the scope delineated in section~\textit{\nameref{subsec:Scope}} is presented in the Supporting Information. They cover basic types of questions that can be asked about processes in which a closed system that contains an ideal gas undergoes a single change of state, leading from an initial equilibrium state to a final equilibrium state. KnowTD solved all these problems successfully. In the Supporting Information, also the solution to the problem is given, together with the reasoning graph that was used to find it.

We also provide a ready-to-use tool that enables the solution of simple thermodynamic problems from the class described above, provided that a solution can be found by single equation traversal. How the tool can be accessed and how it works is described in the Supporting Information. Input files for the 13 problems mentioned above can also be found there. 
We also disclose the entire KnowTD code, including the ontology. How it can be accessed is also described in the Supporting Information. We emphasize that KnowTD is an ongoing project and that this code is only preliminary and not more than a beta version.

\section{Conclusions}
\label{sec:conclusions}

 We have carried out a case study that demonstrates that scientific and engineering knowledge can be transferred to computers and can be made actionable. This was done here for the basic principles of thermodynamics, but similar approaches should be possible for other fields of science and engineering. Major challenges have to be addressed in such an endeavor: firstly, a consistent ontology has to be developed, which has to be modular and open to enable adaptions, extensions, and interfacing, as there are no isolated islands of knowledge. This ontology must not only comprise the pertinent theory of the field but also empirical knowledge of the objects the theory is dealing with, such as material properties. To make the knowledge actionable, the ontology must also contain some information on the problems to be solved. As the ontology contains general knowledge, the key step in applying it to solving a given problem is creating instances. This includes creating instances of the mathematical equations that describe the given problem. Solving these equations is then essentially a mathematical problem. We have focused here on the task of finding a correct system of equations and have discussed only a simple option for solving it, which has a restricted scope as we require that the solution can be obtained by sequentially solving the pertinent equations. Better mathematical routines need to be developed in future work. Thereby, also the question should be addressed how redundant equations should be handled, as the system of equations that we set up contains such redundancy, which makes it different from what would be commonly labeled as a mathematical model of the problem. The result we obtain is not simply the solution in terms of the values of the target variables. Also, the way that solution was obtained, including, e.g., the equations that were used, can be reported. There are many more fascinating options here that we have not yet explored: the system could ask for additional specific input if it recognizes that the equations cannot be solved or if there is ambiguity; the system could also check the consistency of the input and eventually prove that for a given input no solution can exist, to name only some options.
 
 Regarding the thermodynamic knowledge, we have only taken the first steps here. Only a tiny fraction of the thermodynamic knowledge is covered, but the system we have created is open and waits to be extended. Such an extension also has a social dimension: how can and should it be organized? We believe that it should be an open living system. However, its evolution must be organized in a way to maintain the essential feature of the system: the guarantee of correctness. Our next steps are to integrate material property databases into KnowTD and to extend its scope to real fluids, which will greatly enhance its applicability.

\begin{acknowledgement}
We gratefully acknowledge the funding by Deutsche Forschungsgemeinschaft DFG in the frame of the Priority Program 2331 ``Machine Learning in Chemical Engineering''. We also thank Nicolas Hayer for his work on the first prototype of KnowTD as well as Pascal Zittlau and Macro Hoffmann for their work on the examples. 
\end{acknowledgement}

\section{Data and Software Availability}
We provide a ready-to-use implementation of KnowTD, details on the ontology as well as the used sample problems. The prototype of KnowTD is available as a web service at \href{https://knowtd.onrender.com/}{https://knowtd.onrender.com/}. The source code can be found at \href{https://gitlab.rhrk.uni-kl.de/knowtd/knowtd/}{https://gitlab.rhrk.uni-kl.de/knowtd/knowtd/}. Also, the LinkML schemas of the ontology as well as the input files for the sample problems are provided under the above links.

\begin{suppinfo}
\label{suppinfo}

The Supporting Information contains details about the ontology, a description of the prototype as well as sample problems and their solution obtained by KnowTD. It also contains some additional information on the beta verison of the KnowTD code.

\end{suppinfo}
\bibliography{knowTD-bib}

\end{document}